# Earth, a planetary PCR machine to create life, or the brief history of a tRNA

*Juan Jimenez*

Centro Andaluz de Biología del Desarrollo, Universidad Pablo de Olavide/Consejo Superior de Investigaciones Científicas, Carretera de Utrera Km1, 41013 Seville, Spain

Correspondence: jjimmar@upo.es

**Abstract**

About 4 billion years ago, the Earth probably fulfilled the environmental conditions necessary to favour the transition from primitive chemistry to life. Based on a theoretical hairpin duplication origin of tRNAs and their putative peptide-coding capability before ribosomes existed, here I postulate that, at this hypothetical environment, Earth's daily temperature cycles could have provided a unique "planetary thermocycler" to create self-replicating RNA hairpins that simultaneously templated amino acids polymerization in a primordial 'PCR well' of prebiotic molecules. This early RNA hairpin-peptide interaction could have established a reciprocal nucleopeptide replicator that paved the way for catalytic translation and replication machineries towards the origin of LUCA.

## 1 From primitive chemistry to life

The origin of life and of the diversity of the species that inhabit our planet, including ourselves, is an intriguing question that accompanies the history of humans to the present day. Today, from a scientific point of view, the stunning accumulation of genomic sequences in databases and the myriad of bioinformatics tools developed for their in-depth analysis [1-3], make it clear that the fascinating history of life on Earth is written in the pangenome of all living organisms. Comparison of divergent DNA, RNA and/or protein sequences with common ancestors allows us to the theoretical reconstruction of this history back to the first autonomous cell type on our planet that possessed, within a lipid envelope, a nucleic acid synthesis machinery and ribosomes to translate nucleic acids into proteins using a genetic code [4], this hypothetical cell called the Last Universal Common Ancestor LUCA [5-7].

Prior to the origin of life, at a very early stage before self-replicating molecules existed, it is widely accepted that complex organic molecules could be present on Earth that served as building blocks, such as polyphosphate nucleotides and phosphoryl amino acids [8-10], molecules with properties and structures similar to present-day nucleotides and amino acids that could have served as precursors of nucleic acids and proteins in the origin of living systems. These building blocks could have originated by chemical evolution on Earth [11, 12] or provided directly by comets and/or meteorites [13, 14]. But in any case, the key event in prebiotic evolution is how self-replicating nucleotide polymers (nucleic acids) could have originated from monomers. Similarly, how amino acids polymerized to give rise to peptides remains an open question. Moreover, even when nucleic acids and peptides could originate from phosphate transfer activity or other non-specific catalysis [15], it is difficult to envision how independent random polymerization events could generate nucleic acid-coding peptides, and peptides catalysing nucleic acids replication, one of the most mysterious chicken-and-egg riddles in evolution.

Discovering the origin of self-replicating nucleic acids that could simultaneously encode catalytic peptides is the Rosetta stone for understanding the key transition from chemical to biological evolution. Of course, self-replicating RNA molecules encoding peptides or even living protocells could have been 'inoculated' on Earth by comets or meteorites (panspermia theory) [16, 17]; but the existence of different autocatalytic RNAs and ribozymes that catalyze the formation of peptide bonds [18, 19) suggests that RNAs created on Earth (the RNA world) would be a plausible precursor to the much more complex DNA-RNA-protein system on which life today is based [20].

At the chemical to biological transition, many different hypotheses have been proposed [21], but possibly the most widely accepted postulates that the first self-replicating molecule was an RNA strand capable of folding into an active replicase [20-23]. However, the links between self-replicating nucleic acids and the RNA world remain elusive [24-26]. Reciprocal nucleopeptide replicators have also been postulated to play a key role in these early steps of life [27, 28]. Nonetheless, how primordial RNAs capable of self-replication and how encoding catalytic peptides originated is still intriguing. Here, based on the plausible evolutionary history of transfer RNAs (tRNAs), a parsimonious model is presented to answer how self-replicating RNA hairpins and reciprocal nucleopeptide replicators could have been created. Hydrogen-bond-mediated stereochemical interactions between nucleotides and amino acids, subjected to daily temperature fluctuations on Earth, are at the heart of the model.

## 2 Present tRNAs: the clue

The contemporary translation machinery (ribosome) synthesizes proteins encoded in messenger RNAs (mRNAs) and comprises a large set of ribosomal proteins, tRNAs and small and large ribosomal RNAs (rRNAs) that form an active site pocket, likely remnant of the protoribosome [29]. Interestingly, present day rRNAs presumably derived from tRNA sub-elements [30], underlying that tRNAs could be fossil molecules that predate ribosomes [31, 32].

Among the set of protein-coding and non-coding genes found in the genome of all living cells, tRNA sequences represent one of the most abundant and conserved sequences in all living kingdoms [33, 34]. These RNA molecules transfer amino acids to the ribosome for the synthesis of proteins encoded by mRNAs; but canonical tRNAs do not encode proteins. Surprisingly, when virtually translated, a hypothetical peptide derived from an *Escherichia coli* tRNA-Asp sequence is found in the ribosomal protein L27 [35]. This peptide could respond to a randomly conserved fossil sequence in this tRNA. But this observation not only supports that tRNAs might be living fossils that predate ribosomes; it also suggests that tRNA sequences might be at the origin of the translation machinery, encoding the earliest proteins that afterward produced the ribosomes (Fig. 1).

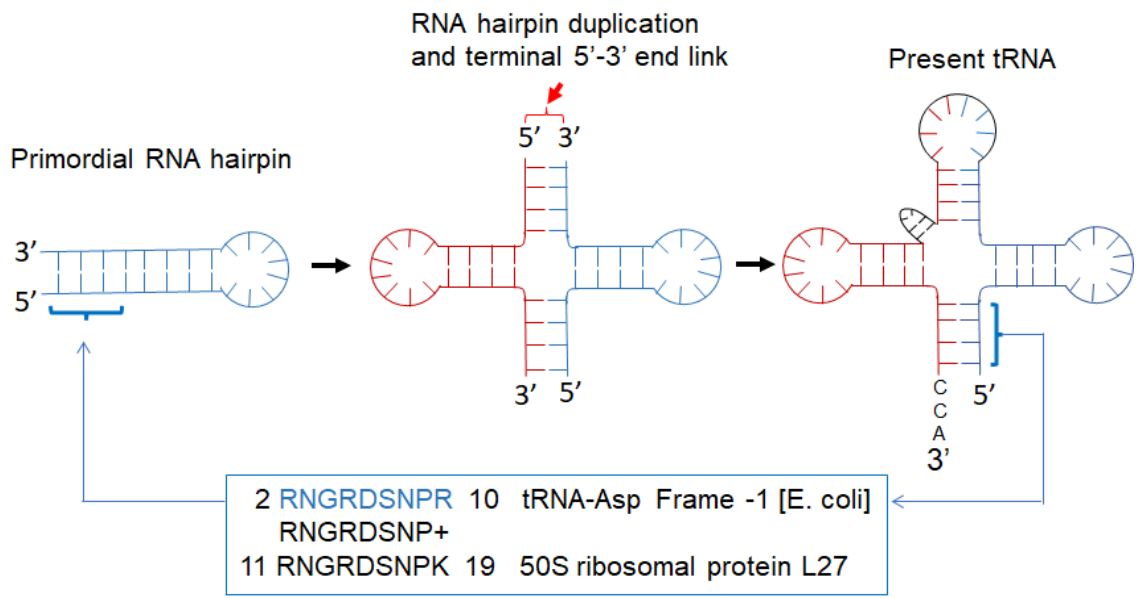

**Fig. 1. Clues to the structure and function of the first molecules of life from present-day tRNAs.** Present day tRNA molecules could have been created from a direct duplication of an ancestral primordial hairpin, with a 5'-3' terminal linkage of the resulting cruciform structure and its subsequent evolution (adapted from [34]). An *E. coli* tRNA-Asp virtually encoding a stretch of the ribosomal protein L27 (square alignment) [35] suggests that tRNAs, and more likely their hairpin precursors, could have encoded peptides before ribosomes existed.

A typical tRNA consists of 72-96 nucleotides with a classic 'cloverleaf' secondary structure that must have originated by direct duplication of a stem-loop hairpin RNA [31, 34] (Fig. 1). Thus, an RNA hairpin could be the simplest structure that could have been created from the polymerization of free phosphonucleotides at the very early origin of life. Therefore, tRNAs are not only representative molecules of the early RNA world. These molecules may also provide direct clues as to how the first self-replicating RNAs were generated and how RNA-encoded proteins could have originated at this early stage of life. The structure and function of tRNA may thus become an invaluable molecular Rosetta stone for understanding the interconnected origin of nucleic acid and protein languages.

## 3 Hydrogen to covalent bonds: the spark that could have initiated life

Stepping back through the evolutionary history of tRNAs, an RNA hairpin could then represent a simple primigenial structure. At this point, the key question is how a single hairpin RNA polymer could have been generated and, even more relevant, how it could have become a self-replicating structure. Assuming that a prebiotic water reservoir enriched with organic chemicals existed on Earth [11, 23], polynucleotides could have been generated by 5'-3' reactions of pre-existing phosphonucleotide monomers in this 'primordial soup'. But even in an environment crowded with molecules, RNA-like polymers should not be produced simply by the 5'-3' binding of individual nucleotides from random Brownian motions. It would be more likely to be caused by physical conditions that could hold two nucleotides together, perhaps by stacked self-associations [36], by some kind of solid template such as clay [37] or rock glasses [38], aided by wet/dry cycling [39], chemical catalysis [15], or activated nucleotides [40], thus lowering the activation energy barrier needed to form phosphodiester bonds between monomers.

Of all the known chemical bonds and forces, only hydrogen bonds can mediate the weak interactions that temporary hold two given molecules together [41]. Notably, a salient property of nucleotides is the ability of their nitrogenous bases to fit together by hydrogen bonding in an antiparallel configuration (adenine with uracil and guanine with cytosine in RNA), a simple but historic discovery by Watson and Crick [42] that marked a milestone in modern molecular biology. Thus, given thermal limits, intrinsic base pairing of nucleotides (or more active primordial analogues [10, 40]) offers a conceptual framework for transient molecular associations to move from a "hydrogen bonding stage" to a "covalent linkage" in prebiotic chemistry [43, 44].

Traditionally, it has been thought that prebiotic chemistry had to have taken place in a rather thermally active environment, the "warm little pond" idea proposed by Darwin. But the binding and polymerization of monomers occurs with high efficiency at low temperatures (around 0°C) [45], a cold scenario that is gaining ground in the RNA community [16, 46, 47]. Spontaneous base pairing of plausible prebiotic nucleotides has been observed in aqueous solutions [30]. Therefore, I speculate that two free nucleotides (probably active analogues [10]) paired by hydrogen bonds at low temperature might have favoured the 5'-3' covalent link between them to produce dinucleotides, a phosphodiester bond that could well be the "spark" that started life some 4 billion years ago in the primordial soup.

# 4 Cyclic thermal fluctuations on Earth: creating self-replicating RNA hairpins

Most studies of the hypothetical life-creating environment have been conducted ignoring daily temperature fluctuations. But nucleotide base pairing is highly dependent on temperature. Hydrogen bonds can be strong enough to bring together matching base pairs at low temperatures, but not when the environmental temperature exceeds a threshold, a remarkable physical property of nucleotides that led to the discovery of PCR, a revolutionary *in vitro* DNA replication technique based on cyclic temperature fluctuations [48]. On Earth today, the temperature in warm deserts can fluctuate from around 0°C at night to near 60°C at midday. It is not known whether the Earth's surface lingered in about 70°C at the Archaean or cooled to freezing conditions in the Hadean [49], but there is strong evidence that in the late Hadean-early Archaean Eon interval when life emerged [50], the Earth's ocean surface was cold but not completely frozen, likely due to high concentrations of greenhouse gases [49-51]. At this period, the first continents were likely formed to support lakes [52, 53]. Therefore, if significant thermal fluctuations could have occurred at the location on Earth where life began, I postulate that daily temperature cycles could then have functioned as a thermocycler, a 'planetary PCR-like machine' in a primordial 'PCR well' soup at this early stage of life (Fig. 2a). Recent experiments have shown that repeated freezing and thawing of aqueous solutions allows the assembly of active ribozymes from inactive precursors encapsulated in separate populations of lipid vesicles [54]. Thus, perhaps night-cooling and day-warming resulting in water freezing-thawing daily cycles could be sufficient at this early Earth to create life.

Thus, the phosphodiester bond described above could have created primordial dinucleotides from hydrogen-bonded nucleotide pairs at low nighttime temperature. During the day, the increased temperature stably linearises the dimer. In a next cycle, the low overnight temperature would have allowed the dimer to assembly two complementary nucleotides that bind 5'-3' to each other, but eventually, also at the 5'-3' terminal end. After daytime heating, a linear four-nucleotide molecule would then have been generated to template at nighttime the assembly and binding of new nucleotides. The process could have continued cycle by cycle, like a PCR machine, producing a double-stranded RNA that, by eventual 5'-3' linkage at one end, gives rise to a terminal loop that doubles the length of the RNA strand (Fig. 2b). Until the length of the RNA stem would have generated a stable stem-loop hairpin molecule that could template the synthesis of complementary, but identical, RNA hairpins on a cycle-by-cycle thermal basis. Thus, as simplified in Fig. 2c, daily cooling and warming cycles of the Earth would have provided the conditions for creating self-multiplying RNA hairpin molecules in a primordial soup, i.e., a lake formed by the impact of a meteorite or comet highly enriched in organic molecules [55], which could have functioned as a PCR well of a planetary PCR-like machine (see in Fig 2a).

Nevertheless, it is difficult to figure out what could have happened in a highly entropic primordial well subject to daily fluctuating temperatures. Even when focusing only on nucleotides, many partial polymerizations randomly initiated in different RNA templates could have yielded an enormous diversity of molecules (see Fig. 2c). But only one hairpin can fully replicate by using itself as a template [23]. Thus, it might have taken many thousands of yeasts to create a self-replicating RNA hairpin, but it makes sense that this "first Darwinian ancestor" could ultimately be the most abundant RNA secondary structure in the prebiotic to biotic era (Fig. 2c).

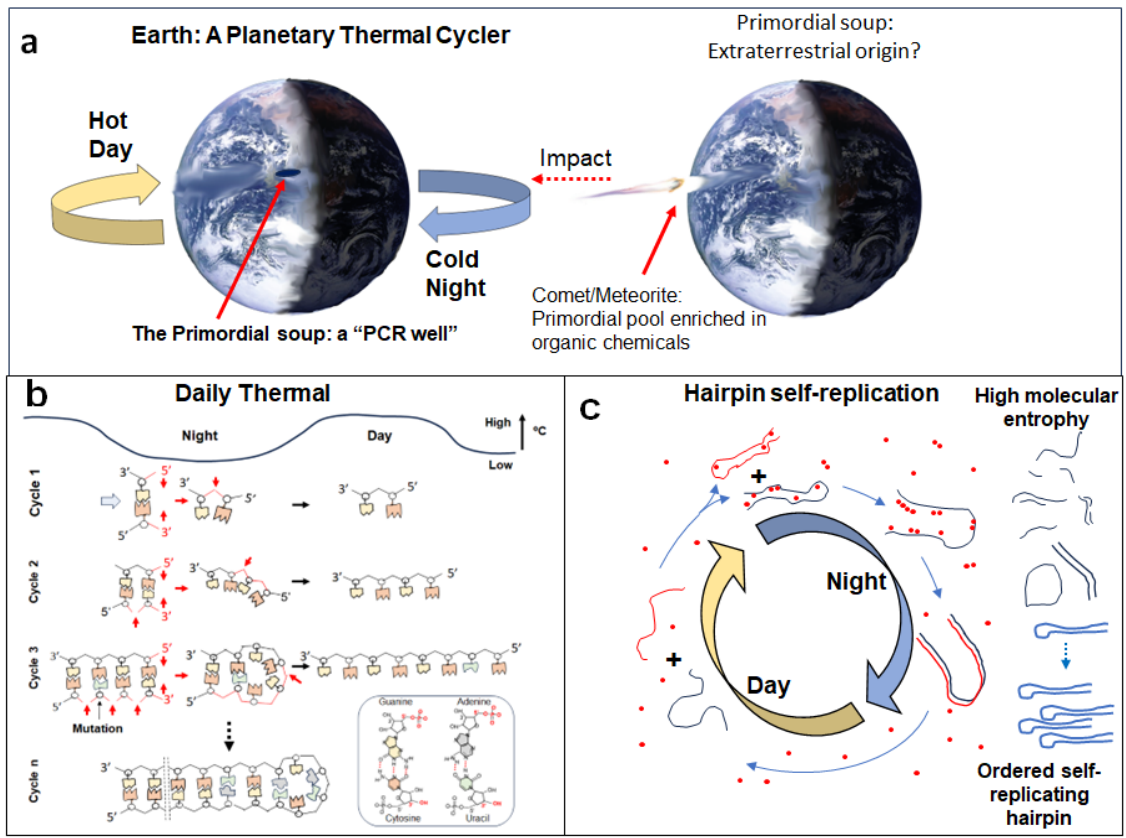

**Fig. 2. The origin of self-replicating RNA polymers.** a) Earth's daily thermal cycles could have functioned as a planetary PCR machine to assist the polymerization of nucleotides into a primordial soup acting as a PCR well (left scheme). This primordial soup could well be of extraterrestrial origin (right scheme). b) The pairing of hydrogen-bonded nitrogenous bases could have favored 5′-3′ covalent bonds of nucleotides at low temperature, including the link of the 5′-3′ terminal ends, which would give rise to a new linearized RNA template at high temperature, replicating the produced RNA molecules cycle by cycle. c) This highly entropic reaction would produce a great diversity of molecules (right scheme), but only one hairpin can fully replicate using itself as a template, and this molecule is expected to reach a long-term steady state of self-replicating molecules (left scheme).

## 5 Stereochemical nucleotides-amino acids associations: RNA-templated synthesis of peptides

As for free nucleotides, it has been proposed that spontaneous polymerization of free amino acids could have originated peptides in the early stages of life [20, 56]. But from independent nucleotide and amino acid polymerization events, it is extremely difficult to sustain a plausible convergence of both random polymers into interconnected RNA replication and translation machineries. Since tRNAs are part of the protein synthesis machinery but likely predate ribosomes, protein synthesis must have been originated in tRNA-related structures [35]. Interestingly, a tRNA-like molecule covalently linked to a polypeptide has been proposed as an intermediate step between the RNA word and the origin of protein synthesis [57, 58]. However, the hairpin structure may represent the probable precursor of the tRNA molecule [25]. Therefore, a first step in the origin of protein synthesis must have taken place in these hairpin structures (see Fig. 1).

One of the first theories to address this difficult question is the stereochemical hypothesis, which postulates that the genetic code evolved from interactions between oligonucleotides containing codons or anticodons and their respective amino acids [59-63]. Interestingly, amino acids in ribosomal proteins associate with ribosomal RNAs in trinucleotides that match the genetic code, a possible fingerprint of this ancestral association [64]. Furthermore, amino acids can interact directly with specific nucleotides [65-67], with dinucleotides being one of the preferred association geometries [61, 68]. Thus, if similar stereochemical associations occurred between primordial amino acids and early hairpin RNA molecules, the specific association of amino acids with their respective "codons" (non-overlapping trinucleotides, about 1 nm in length each) could have brought their amino and carboxyl groups close enough to make peptide bond formation possible [69] (in contemporary proteins, alpha carbons of two peptidyl-linked amino acids are separated by 0.38 nm [70]). This covalent peptide bond formation could have been assisted by the nascent ribozyme activity of existing RNA hairpins [71]. Consequently, early RNA hairpins could have served as templates for both complementary nucleotides (self-replication) and trinucleotide-associated amino acids (prototranslation), giving rise to primordial "encoded" peptides (see Fig. 3a).

## 6 Toward a reciprocal translation-replication machinery

Selective association of amino acids on hairpin RNA templates could have provided a very simple but reliable primordial translation machinery. However, random binding of nucleotides and amino acids on a common template (ignoring other interacting molecules) would be expected to interfere with their respective polymerization processes (Fig. 3b). Interestingly, some amino acids are known to physically fit into an amino to carboxyl configuration at their 5' to 3' trinucleotide codon [60, 62, 64, 67]. Therefore, under the assumption that present-day translation might be a faithful mirror of its evolutionary origins, it is conceivable that hairpin RNA sequences preferentially associating amino acids on the 5' terminal strand of the stem might have increased their polymerization efficiency. This strategy of amino-to-carboxyl synthesis in the 5' to 3' direction of the RNA template might have established the mode of action of present day ribosomes on mRNAs. Interestingly, preferential binding of amino acids to the 5' to 3' strand of the stem would release the 3' to 5' complementary one for nucleotide association, favouring semi-conservative polymerization/replication of the new RNA hairpin in the 5' to 3' direction (see Fig. 3b), as occurs in all living replication/transcription systems.

During the overnight low-temperature amino acid polymerization step, peptides are likely to fold into tertiary structures as they are synthesized [72]. Therefore, it might be expected that upon reaching a critical size, the synthesized peptide would fold to self-release from the RNA template, allowing for the complete synthesis of the new RNA hairpin initiated at the complementary stand (see schematic representation in Fig. 3a). Therefore, rather than competing, preferential association of amino acids to the 5'-end strand of the RNA hairpin template might have introduced order into the entropic environmental system to create efficient "PCR-like" nucleoprotein replicators (Fig. 3c), an important scenario that could explain basic properties of the current replication/transcription (3' to 5' templated) and translation (5' to 3' encoded) machineries.

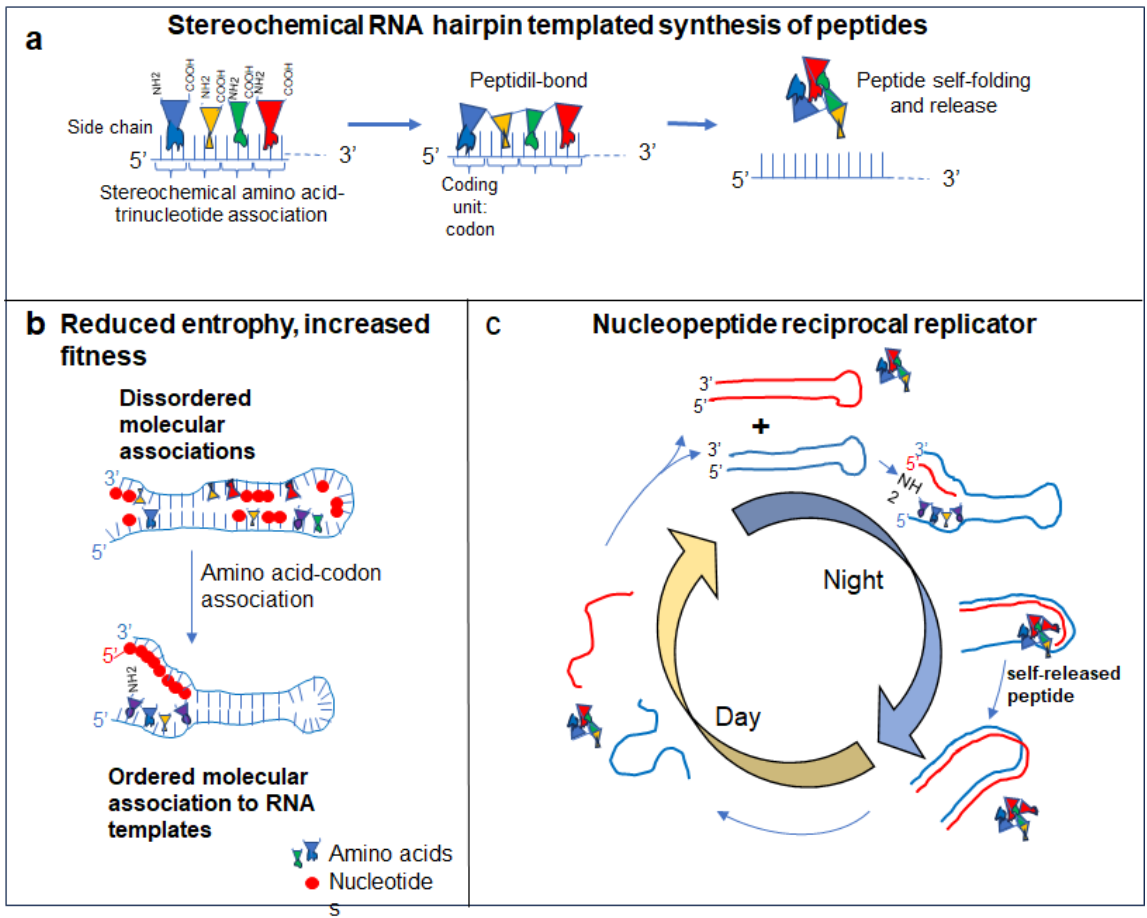

**Fig. 3. Primordial RNA hairpins may have templated the polymerization of nucleotides and amino acids.** a) Stereochemical binding of amino acids to specific trinucleotides (codons) holds together two amino acids in consecutive non-overlapping codons (left scheme) to allow peptidyl binding (middle scheme). The property of peptides to fold into tertiary structures would release the synthesized peptide from the RNA template when reaching a critical size (right scheme). b) Amino acids and nucleotides would be expected to compete for binding to a common RNA template (top diagram). Darwinian selection of the ordered assembly of amino acids to the 5'-terminal strand, releasing the 3' terminal one for nucleotide binding, could have significantly increased the efficiency of both peptide synthesis and RNA hairpin self-replication (bottom diagram). c) Selective association of small amino acids to the 5' strand could have led to an early nucleopeptide reciprocal replicator. Synthesized peptides that assist hairpin replication could have been precursors of the nucleic acid replication machinery itself.

Based on this hypothesis, I speculate that random mutations in the primordial RNA hairpins that eventually ordered polymerization processes into separate strands would have been rapidly selected, leading to an important step in the evolution of life.

## 7 Amino acid side chain diversity: the need for transfer RNA hairpins

In the stereochemical (preribosomal) synthesis of peptides discussed above, the direct association of amino acids to codons must be discriminated by the binding properties of their variable side chains with the Watson-Crick side of the nucleotide bases [65]. Amino acid side chain lengths can vary from 0.4 to 1 nm. Therefore, it could happen that when adjacent amino acids have very different side chains, their respective amino and carboxyl groups may be physically too far apart when associating with their codons on the RNA hairpin template (primordial mRNA). This physical fact would likely have restricted the formation of peptidyl bonds (Fig. 4a). In agreement with this idea, it has been reported that, regardless of prebiotic abundance, early encoded peptides could have been formed by a reduced pool of small amino acids [73].

In random interactions, hairpins with a single amino acid associated to a 5' end codon could have been transiently originated. In these single amino acid-hairpin RNA conjugations, codon recognition by the amino acid can be transferred to the "anticodon", the triplet complementary to the codon located at the 3' end of the stem, thus converting this amino acid-RNA hairpin conjugation into a "transfer RNA hairpin" (tRNA precursors) (Fig. 4b), an oligonucleotide adaptor predicted by Crick [74]. Interestingly, nucleotide sequence analysis of 1400 tRNAs revealed the imprint of a prototypical genetic code of their corresponding amino acids at position 3-4-5 (5' end) [75]. Thus, by mediating in codon recognition, the transfer hairpin could closer together amino and carboxyl groups of adjacent encoded amino acids, thus overcoming the physical limitations imposed by their length differences (Fig. 4).

Transfer RNA hairpins could have evolved into tRNAs by a duplication [22, 25] (Fig. 1), in which the anticodon of the 3'-terminal end of the stem is transferred to an opposite anticodon loop, probably optimizing COOH-NH2 binding of the growing peptide and codon-anticodon recognition (including wobbling at the third codon position [76]) (Fig. 4c). The amino acid could have been covalently transferred from the 5' end (codon-associated) to the 3' end of the acceptor stem where localize in present tRNAs [71] (Fig. 4c), opening the possibility of incorporating new amino acids (i.e., those associated with anticodon sequences) to complete the genetic code [74 , 71].

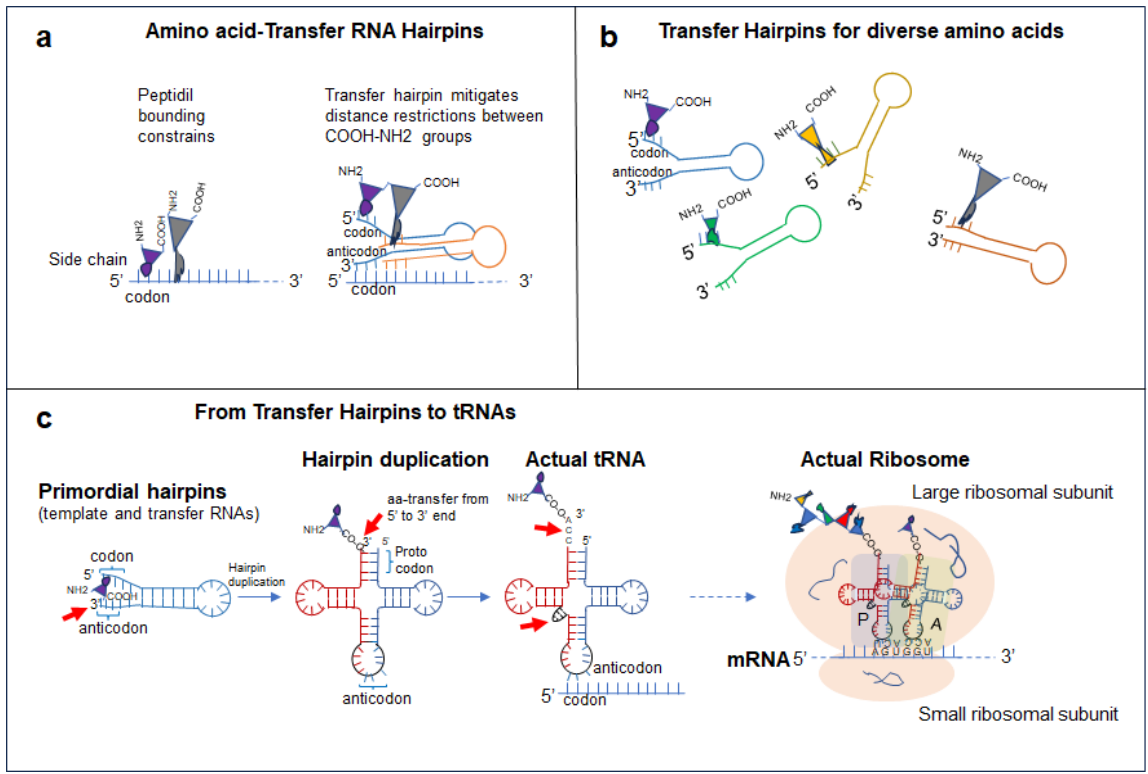

**Fig. 4. Transfer RNA hairpins could have enhanced peptide synthesis in primordial RNA templates.** a) Following stereochemical association to codon triplets, different side chains could constrain peptide binding of adjacent amino acids (left scheme). Single amino acid-codon hairpin conjugates could transfer codon recognition to the complementary anticodon of the RNA hairpin, thus buffering side-chain differences (right scheme). b) Transfer RNA hairpins could have diversified to associate specific amino acids. c) Evolutionary steps from simple transfer RNA hairpins to contemporary tRNAs (key steps highlighted by red arrows). Diverse RNA hairpins acting as a template (proto mRNA) and as an amino acid transfer (proto tRNA) could have produced proteins that aggregated to produce highly efficient translation machinery (present ribosomes).

Overall, these evolutionary steps towards contemporary tRNAs have likely been very dynamic [77], standardizing the COOH-NH2 proximity of amino acids encoded in adjacent codons on mRNA templates to enhance peptidyl-binding efficiency and translation accuracy.

## 8 Concluding Remarks

The Earth originated 4.5 billion years ago and about 500 million years later developed conditions suitable for supporting life [49]. It has recently been inferred that LUCA lived 4.0–4.3 billion years ago [7]. Thus, although the chemical to biological transition would be expected to take billions of years in fixed environments, life could originate quite rapidly on geological timescales. The "PCR-like" mode of action of Earth postulated here may cast some light on this paradox.

In primordial RNA hairpins created by a "planetary thermal cycler" multiplication from hydrogen-paired nucleotides, the replication step should be error-prone through frequent non-Watson and Crick base pairings [78] (see in Fig. 2b), yielding high RNA sequence diversity. Since mutations in the RNA sequence could change its "heritable" stability, replication and/or prototranslation properties, individual stem-loop hairpin molecules might have been subject to fitness selection in a Darwinian "generation-by-generation" manner [16], cycle-by-cycle in a primordial soup on Earth that functioned as a planetary PCR-like machine.

The evolution of the sequence encoding peptides that enhance RNA polymerization might have eventually led to proto-RNA polymerases (see in Fig. 3c). At this stage, RNA self-replication and peptide formation might have become progressively less dependent on temperature cycles and more dependent on their direct peptide-RNA interactions and catalytic activities. Since proteins can function in trans (binding to diversified RNAs, including rRNAs), RNA hairpin units encoding peptides capable of forming peptide-RNA aggregates could have been rapidly selected, resembling in some aspects existing ribosomal aggregates [29] (see Fig. 4c). Furthermore, duplication of RNA hairpins and/or cross-hybridization could have generated single RNA molecules with several different hairpins (protogenes?), mimicking primordial chromosomes that differentiated replication (full-length copy) from transcription (partial RNA copy covering one or a few hairpin units).

As organic compounds were depleted in the original pool, molecules with similar physicochemical properties could have replaced these primordial building blocks with currently existing nucleotides and amino acids, some specialized in hosting genetic information (DNA in living cells) and others in protein synthesis (mRNA, rRNA and tRNA) yielding a large repertoire of catalytic and structural functions. This repertoire of encoded proteins should have given rise to metabolic pathways capable of providing the molecules that were progressively depleted in the primordial environment [79]. The eventual emergence of proteins enriched with hydrophobic amino acids may have favoured non-covalent lipid conjugation to enclose these proto-replication and proto-translation machineries, thus enhancing fitness and cooperation within the aggregates. And from here, an amoeboid-type division [80] and a basic regulation of the cell cycle [81] possibly cleared the way to the origin of the first LUCA cell, capable of propagating life outside the primordial "PCR well" of the Earth.

## Acknowledgments

I thank all attendants to the lab retreat at Paul Nurse's honorary doctorate at UPO for their helpful comments and discussions. This work was supported by the Spanish Ministerio de Ciencia e Innovación (grant number PID2019-111124GB-I00).

## Contributions

I, Juan Jimenez, am the sole author.

## Declaration of interests

The author declares no competing interests.